\documentclass{article}
\usepackage{geometry}                
\usepackage{graphicx}
\usepackage{amssymb}
\usepackage{epstopdf}
\usepackage{latexsym}
\usepackage{dcolumn}
\usepackage{epsfig}
\title{Ultrahigh energy neutrino scattering: an update}
\author{{Martin~M.~Block}\\
 {\it Department of Physics and Astronomy, Northwestern University, Evanston, IL 60208}
\and
Phuoc Ha\\
{\it Department of Physics, Astronomy and Geosciences, Towson University, Towson, MD 21252}
\and
Douglas W. McKay\\
 {\it Department of Physics and Astronomy, University of Kansas, Lawrence, KS 66045}}
\date{\today}

\begin{document}
\maketitle

\begin{abstract}
We update our estimates of charged and neutral current neutrino total cross sections on isoscalar nucleons at ultrahigh energies using a global (x, $Q^2$) fit, motivated by the Froissart bound \cite{froissart},  to the $F_2$ (electron-proton) structure function utilizing the most recent analysis \cite{hera} of the complete ZEUS and H1 data sets from HERA I. Using the large $Q^2$, small Bjorken-$x$ limits of the ``wee'' parton model, we  connect the ultrahigh energy neutrino cross sections {\em directly} to the large $Q^2$,  small $x$  extrapolation of our new fit, which we assume saturates the Froissart bound \cite{froissart}.  We compare both to our previous work \cite{bbmt}, which utilized only  the  smaller ZEUS \cite{zeus01} data set, as well as to recent results \cite {c-ss} of a calculation using the ZEUS-S based global perturbative QCD parton distributions using the combined HERA I results as input.  Our new results substantiate our previous conclusions \cite{bbmt}, again predicting { \em significantly  smaller} cross sections  than those predicted by  extrapolating pQCD calculations to neutrino energies above $10^9$ GeV.
\end{abstract}
\section{Introduction}

Large experiments seeking  evidence of ultrahigh energy (UHE) cosmic neutrinos have been going on in earnest for more than a decade.\footnote{For purposes of this paper, we take UHE to mean $E_{\nu} \geq 10^6$ GeV and cosmic to mean origins that are galactic, extragalactic or cosmogenic.} Searches have been underway with detectors scanning for UHE cosmic neutrino induced events in large volumes of water \cite{baikal, antares}, ice \cite{icecube, rice, forte, anita}, Earth's atmosphere \cite{HiRes, Auger} and the lunar regolith \cite{glue}.  No clear indication of a cosmic neutrino event has yet been reported, but by assuming extrapolations of neutrino-nucleon cross sections from low energy data to UHE, the experiments have all reported bounds on UHE neutrino flux models.  Put together, the experimental bounds on neutrino fluxes now cover energies up to $10^{17}$ GeV.  

Given the importance of neutrino cross section estimates in the process of interpreting data (i.e., placing flux bounds) or planning new experiments (i.e., EUSO \cite{euso}), having a range of realistic predictions of neutrino cross sections available is crucial to the ultimate success of the  UHE neutrino physics enterprise.  It was in this spirit we applied an extrapolation to UHE of a global fit to $F_2$ structure function data \cite{bbt} from ZEUS  \cite{zeus01} to an estimate of the UHE neutrino-nucleon total cross section \cite{bbmt}.  The fit function was well motivated, invoking the Froissart $\ln^2(s)$ asymptotic bound on hadronic cross sections \cite{froissart}, \cite{bh}, and with only a handful of parameters, it provided an excellent fit to the ZEUS data set which covered an impressive range of $Q^2$ and $x$ values.  Though the agreement with the conventional perturbative quantum chromodynamics (pQCD) prediction was good in the range of $E_{\nu}$ where the cross section is dominated by the HERA $(x, Q^2)$ domain, when extrapolated to energies above
$10^9$ GeV our prediction fell well below the pQCD extrapolation, having important implications for the estimate of event rates for the highest energy experiments.  

With the significance of these issues in mind, we devote this paper to updating the results of our earlier calculation by using the recently published joint ZEUS and H1 analysis of \emph{all} the HERA I data \cite{hera}. In particular, this new analysis reconciles some tension between relative normalization between the two experiments, making a complete evaluation of errors for the combined data set.  We fit the new results as before, recalculate $\sigma_{\nu N}(E_{\nu})$ and compare both with our previous results \cite{bbmt}, as well as new results of a standard pQCD calculation based on a recent determination of quark distribution functions \cite{c-ss} using a earlier analysis \cite{zeus03}, updated to include all the HERA I data.  We confirm our previous conclusion that our unitarity bounded cross sections predict values that are significantly lower than those found from unscreened pQCD at neutrino energies above $10^9$ GeV. Moreover, even when we extrapolate our fits many orders of magnitude in energy above the region covered by the HERA data, the greatly expanded and re-analyzed data base leads to only very modest changes in our new values for $\sigma_{\nu N}$ when compared to our previous ones \cite{bbmt}.

\section{The UHE $\nu -N$ cross section from the full HERA I data set}

We assume two full families of active quarks. In the ``wee parton'' picture, assuming that valence quark contributions are small at UHE and that sea quarks all contribute the same (equipartition of flavors), the neutrino-isoscalar nucleon N (N=(p+n)/2)  double differential cross section can be written as 

\begin{equation}
\frac{d^2\sigma_V}{dxdy}(E_{\nu})=\frac{2G_F^2mE_{\nu}}{\pi}\left(\frac{M_{V}^2}{Q^2+M_{V}^2}\right)^2\times\left[\kappa_V F_2(x,Q^2)\right](1+(1-y)^2),
\label{masterEq}
\end{equation}
where $E_{\nu}$ is the neutrino energy in the rest frame of the nucleon.
The electron-proton structure function $F_2(x,Q^2)$ is provided by our fit to the combined ZEUS and H1 analysis of the full HERA I data set\cite{hera}, described in the following section. The label V takes the values V = CC (charged current) and NC (neutral current), where $M_{CC}$=$M_W$, $M_{NC}$=$M_Z$, $m$ is the mass of the nucleon, $G_F$ is the Fermi weak coupling constant, and 
$\kappa_{CC}$=$\frac{9}{10}$ and $\kappa_{NC}$=0.298 for two full families, using the 2008 PDG  \cite{pdg} value of $\sin^2(\theta_W)$ at $Q^2$=$M_Z^2$ in the $\overline{MS}$ scheme. 
The factors relating the  electron-proton $F_2$ structure function (found in deep inelastic scattering) to the CC and NC equivalents, using the ```wee parton'' model,  are derived in \cite{bbmt}. The general quark parton CC and NC expressions are given in \cite{c-ss,gqrs98}.  

\subsection{Kinematics}

In  Eq.(\ref{masterEq}), the inclusive double differential cross section for the CC (NC) neutrino inclusive cross sections for the processes $\nu_{\ell}(k)+N(p) \rightarrow \ell(k')+X_{CC}$  $(\nu_{\ell}(k')+X_{NC})$, $\ell$ = e, $\mu$, $\tau$, depends upon the invariants $s=(k+p)^2$, $-Q^2=(k-k')^2$, and $p\cdot (k-k')$.  The scaling variables are $x=Q^2/2p\cdot(k-k')$ and $y=p\cdot(k-k')/p\cdot k$.  If $E_{\nu}\gg m$, $s=2mE_{\nu}$ and $Q^2=2mE_{\nu}xy$. In  the rest frame of the nucleon, $y= 1-E_{\nu}'/E_{\nu}$, the fractional energy transferred from the incoming neutrino to the outgoing lepton or neutrino. 

 In the integrations over $x$ and $y$, whose results we will report below, we set a limit to the minimum value of $Q^2$, which avoids possible numerical problems with singular behavior of the integrands as $x$ or $y$ goes to zero.  Given the relationship shown in the preceding paragraph, we see that this amounts to limiting $x$ and $y$ to the ranges $Q^2_{min}/(2mE_{\nu})\leq x\leq 1$ and, taking $y$ as the initial integration, $Q_{min}^2/(2mE_{\nu}x)\leq y\leq 1$.
 
 As recognized early in \cite{mr85}, the vector boson propagator factor ($M_{V}^2/(M_{V}^2+Q^2))^2$ acts to cut off the integrand for $Q^2 >M_{V}^2$, effectively selecting a range of small $x$---reaching somewhat below $x\sim M_{V}^2/(2mE_{\nu})$---that makes substantial contributions to the total cross section.  For the range of neutrino energies we consider in this work,  $10^4$ GeV $\leq E_{\nu}\leq 10^{14}$ GeV, this means that we probe $x$ values in the range $0.1\leq x\leq10^{-11}$!

\subsection{Global fit to combined ZEUS and H1 $F_2(x,Q^2)$ results}

 We follow the work of Berger, Block and Tan \cite{bbt}, who obtained a remarkably good fit to ZEUS small $x$ results for $F_2(x,Q^2)$ \cite{zeus01}, using a 6 parameter model guided by  empirical analysis \cite{bh} of experimental data  on high energy hadronic cross sections that empirically demonstrated the saturation of the $\ln^2 s$  Froissart bound \cite{froissart}. They assumed that Deep Inelastic Scattering (DIS) was $\gamma^*p$ hadronic scattering and applied the $\ln^2s$  Froissart bound saturation successfully to the $F_2(x,Q^2)$ data \cite{zeus01}.  For $Q^2\gg m^2$, this bound in $s$ translates into  a $\ln^2(1/x)$ bound on the small $x$ behavior of $F_2(x,Q^2)$. We here apply the same fit function to the recently reported results from a combined ZEUS and H1 determination of DIS $e^{\pm}p$ cross sections \cite{hera}, which provides very accurate values of $F_2(x,Q^2)$ over a large region of the $(x,Q^2)$ plane.  Using Eq. (\ref{masterEq}), we then predict UHE neutrino cross sections based on the extrapolation of the analytic expression $F_2$ into the $Q^2 > M_Z^2$ and $x<10^{-10}$ regions needed to evaluate the cross sections up to $10^{12}$ GeV and beyond.
 
 The global fit function takes the form, for small $x$,
 \begin{eqnarray}
 F_2(x,Q^2) &=& (1-x)(\frac{F_P}{1-x_P}+A(Q^2)\ln[\frac{x_P}{x}\frac{1-x}{1-x_P}] \nonumber \\
                    &  & \mbox +B(Q^2)\ln^2[\frac{x_P}{x}\frac{1-x}{1-x_P}]),
 \label{eqnF2smx}
 \end{eqnarray}
 where
 \begin{eqnarray}
 A(Q^2) &=& a_0+a_1\ln Q^2+a_2\ln^2Q^2 \nonumber \\
 B(Q^2) &=& b_0+b_1\ln Q^2+b_2\ln^2Q^2.
 \label{eqnABsmx}
 \end{eqnarray}
 A $\chi^2$ minimization global fit of Eq.(\ref{eqnF2smx}) and  Eq.(\ref{eqnABsmx}) was made to all of the HERA combined experimental $F_2$ data in both $Q^2$ and $x$, with $x\leq x_P=0.11$.   At the point $x_P=0.11$, $dF_2(x_P,Q^2)/dQ^2$=0 for all $Q^2$. This procedure differs slightly from that in \cite{bbt}, where $F_P$ was fixed at the value 0.41; as in \cite{bbt}, we again used the ``Sieve'' algorithm \cite{mbnim}  to eliminate  ``outlier'' datum points, using a $\Delta\chi^2_{\rm max}=6$ cut.   $F_P$, the value of $F_2$ at $x_P$, 
 along with the other 6 parameters, together with their errors, are listed in Table 1.  Also shown are the renormalized minimized $\chi^2$ value \cite{mbnim}, the number of degrees of freedom and the renormalized $\chi^2$ per degree of freedom for our new analytic form for the combined ZEUS and H1 results \cite{hera}.

The large $x\sim1$ region, which contributes very little the UHE cross sections,  is fitted by the form
 \begin{table}[ht]                   
%
\begin{center}
\def\arraystretch{1.2}            
     \caption{\label{fitted}\protect\small Results of a 7-parameter fit to the
HERA combined data for $F_2(x,Q^2)$ for $0.85 \le
Q^2\le 3000$ GeV$^2$ and $x\le 0.1$. The $\chi^2_{\rm min}$ is renormalized  by the factor $\mathcal{R}$ to take into account the effects of the cut at $\Delta \chi^2_{i \rm max}$ = 6 \cite{mbnim}. \label{table:results}}
\begin{tabular}[b]{|l||l||c||}
    \cline{1-2}
Parameters&Values\\
\hline
      $a_0$&$-8.471\times 10^{-2}\pm 2.62\times 10^{-3}$ \\
      $a_1$&$\phantom{-}4.190\times 10^{-2}\pm 1.56\times 10^{-3}$\\
      $a_2$&$-3.976\times 10^{-3}\pm 2.13\times 10^{-4}$\\
\hline
    $b_0$ &$\phantom{-}1.292\times 10^{-2}\pm 3.62\times 10^{-4}$\\
      $b_1$&$\phantom{-}2.473\times 10^{-4}\pm 2.46\times 10^{-4}$\\
      $b_2$&$\phantom{-}1.642\times 10^{-3}\pm 5.52\times 10^{-5}$ \\
\hline
$F_P$&\phantom{---}$0.413\pm0.003$\\
    \cline{1-2}
        \hline
    \hline
    $\chi^2_{\rm min}$&\phantom{---}352.751\\
    $\mathcal{R}\times\chi^2_{\rm min}$&\phantom{---}391.377\\
    
    d.f.&\phantom{---}335\\
    $\mathcal{R}\times\chi^2_{\rm min}/$d.f.&\phantom{---}1.17\\
\hline
\end{tabular}
\end{center}
\label{tabF2smx}
\end{table}
\def\arraystretch{1}  

 \begin{equation}
 F_2(x,Q^2)=F_P\left(\frac{x}{x_P}\right)^{\rho(Q^2)}\left(\frac{1-x}{1-x_P}\right)^3.
 \label{eqnF2lgx}
 \end{equation}
Equations (\ref{eqnF2smx}) and (\ref{eqnF2lgx}) obviously guarantee continuity of $F_2(x,Q^2)$ at $x=x_P$ for all $x$.  Requiring that $dF_2(x,Q^2)/dx$ is also continuous at $x=x_P$ determines $\rho(Q^2)$ in terms of $x_P, F_P$ and $A(Q^2)$, Eq. (\ref{eqnABsmx}). 
 
\section{Total charged and neutral current neutrino-nucleon cross sections}
The total deep inelastic scattering neutrino-nucleon cross section is obtained by the double integral of Eq.(\ref{masterEq}) over $Q_{min}^2/(x 2mE_{\nu})\leq y \leq 1$ and $Q_{min}^2/( 2mE_{\nu})\leq x \leq 1$.  
Tables 2 and 3, calculated using $Q_{min}^2$ = 0.01 GeV$^2$,  show the charged and neutral current total $\nu$-isoscalar nucleon cross section values in cm$^2$, at energies from $10^4$ to $10^{14}$ GeV. The column labels $\sigma_{HERA}$, $\sigma_{BBMT}$, $\sigma_{C-SS}$ and $\sigma_{GQRS}$ refer, respectively, to a) the calculations based on our new fit-function---Eqs. (\ref{eqnF2smx}) and (\ref{eqnF2lgx})---to the combined ZEUS and H1 measurements of $F_2$ \cite{hera}, b) the published Berger, Block, McKay and Tan (BBMT) values based on the same fit function applied to the older ZEUS $F_2$ values  \cite{bbmt}, c) a recent calculation \cite{c-ss} using ZEUS-S \cite{zeus03} parton distribution function (pdf) fits with NLO pQCD evolution \cite{dglap}, and d) the results from the 1998 paper of Gandhi, Quigg, Reno and Sarcevic \cite{gqrs98} using CTEQ4 fits with DIS pQCD evolution. 

\begin{table}[ht]                   
%
\begin{center}
     \caption{Charged current cross sections, in cm$^2$, as a function of $E_\nu$, the laboratory energy in GeV;  $\sigma_{HERA}$, from our model using a fit to new HERA results;   $\sigma_{BBMT}$, from our earlier fit to ZEUS data \cite{bbmt};  $\sigma_{C-SS}$,  from a standard evaluation from recent pdfs  \cite{c-ss} ; $\sigma_{GQRS}$,  from the 1998, CTEQ4 based pdf \cite{gqrs98}. Note that the latter two only evaluate values  up to $E_\nu=10^{12}$ GeV.} 
   \begin{tabular}{|l|l|l|l|l|r|} 
 \hline 

      $E_\nu$ (GeV)    & $\sigma_{HERA}({\rm cm^2})$ & $\sigma_{BBMT}({\rm cm^2})$  & $\sigma_{C-SS}({\rm cm^2})$ & $\sigma_{GQRS}({\rm cm^2})$ \\  \hline
      
           $10^4$    & 4.57 $10^{-35}$ & 3.80 $10^{-35}$ & 4.50 $10^{-35}$ & 4.62 $10^{-35}$ \\
                
           $10^5$    & 2.11 $10^{-34}$ & 1.91 $10^{-34}$ & 1.95 $10^{-34}$ &  2.02 $10^{-34}$ \\
      
           $10^6$    & 6.88 $10^{-34}$ & 6.87 $10^{-34}$ & 6.01 $10^{-34}$ &  6.34 $10^{-34}$ \\  
           
            $10^7$    & 1.90 $10^{-33}$ & 1.94 $10^{-33}$ & 1.60 $10^{-33}$ & 1.75 $10^{-33}$ \\  
            
            $10^8$    & 4.48 $10^{-33}$ & 4.49 $10^{-33}$ & 3.87 $10^{-33}$ & 4.44 $10^{-33}$ \\  
            
            $10^9$    & 9.09 $10^{-33}$ & 8.90 $10^{-33}$ & 8.76 $10^{-33}$ & 1.05 $10^{-32}$ \\  
            
            $10^{10}$  & 1.64 $10^{-32}$ & 1.58 $10^{-32}$ & 1.87 $10^{-32}$ & 2.38 $10^{-32}$ \\ 
            
            $10^{11}$   & 2.72 $10^{-32}$ & 2.57 $10^{-32}$ & 3.81 $10^{-32}$ & 5.36 $10^{-32}$ \\ 
            
            $10^{12}$   & 4.21 $10^{-32}$ & 3.92 $10^{-32}$ & 7.40 $10^{-32}$ & 1.18 $10^{-31}$ \\   
            
            $10^{13}$   & 6.17 $10^{-32}$ & 5.68 $10^{-32}$ & ... & ... \\ 
            
            $10^{14}$   & 8.68 $10^{-32}$ & 7.92 $10^{-32}$ & ... & ... \\  \hline  
            
        \end{tabular}
        \end{center}
\label{CCx-secs}
\end{table}
      
 \begin{table}[ht]                   
%
\begin{center}
    \caption{Neutral current cross sections, in cm$^2$, as a function of $E_\nu$, the laboratory energy in GeV; $\sigma_{HERA}$, from our model using a fit to new HERA results;  $\sigma_{BBMT}$, from our earlier fit to ZEUS data, \cite{bbmt}; $\sigma_{GQRS}$,  from the 1998 CTEQ4 based pQCD extrapolation of Ref.  \cite{gqrs98}, which quotes values only up to $10^{12}$ GeV.  Note that Ref. \cite{c-ss} does not report NC results. }
   \begin{tabular}{|l|l|l|l|l|r|} 
   
\hline

      $E_\nu$ (GeV)    & $\sigma_{HERA}({\rm cm}^2)$ & $\sigma_{BBMT}({\rm cm}^2)$   & $\sigma_{GQRS}({\rm cm}^2)$ \\  \hline
      
           $10^4$    & 1.60 $10^{-35}$ & 1.32 $10^{-35}$ & 1.58 $10^{-35}$ \\
                
           $10^5$    & 7.88 $10^{-35}$ & 7.03 $10^{-35}$ & 7.67 $10^{-35}$ \\
      
           $10^6$    & 2.65 $10^{-34}$ & 2.65 $10^{-34}$ & 2.60 $10^{-34}$ \\  
           
            $10^7$    & 7.55 $10^{-34}$ & 7.74 $10^{-34}$ & 7.48 $10^{-34}$ \\  
            
            $10^8$    & 1.82 $10^{-33}$ & 1.83 $10^{-33}$ & 1.94 $10^{-33}$ \\  
            
            $10^9$    & 3.76 $10^{-33}$ & 3.70 $10^{-33}$ & 4.64 $10^{-33}$ \\  
            
            $10^{10}$  & 6.89 $10^{-33}$ & 6.63 $10^{-33}$ & 1.07 $10^{-32}$ \\ 
            
            $10^{11}$   & 1.15 $10^{-32}$ & 1.09 $10^{-32}$ & 2.38 $10^{-32}$ \\ 
            
            $10^{12}$   & 1.79 $10^{-32}$ & 1.67 $10^{-32}$ & 5.20 $10^{-32}$ \\   
            
            $10^{13}$   & 2.64 $10^{-32}$ & 2.44 $10^{-32}$ &  ...\\ 
            
            $10^{14}$   & 3.73 $10^{-32}$ & 3.40 $10^{-32}$ &  ...\\  \hline  
            
        \end{tabular}
      \end{center}
\label{NCx-secs}
\end{table}
      
\begin{figure}[htbp] 
   \centering
   \includegraphics[width=4in]{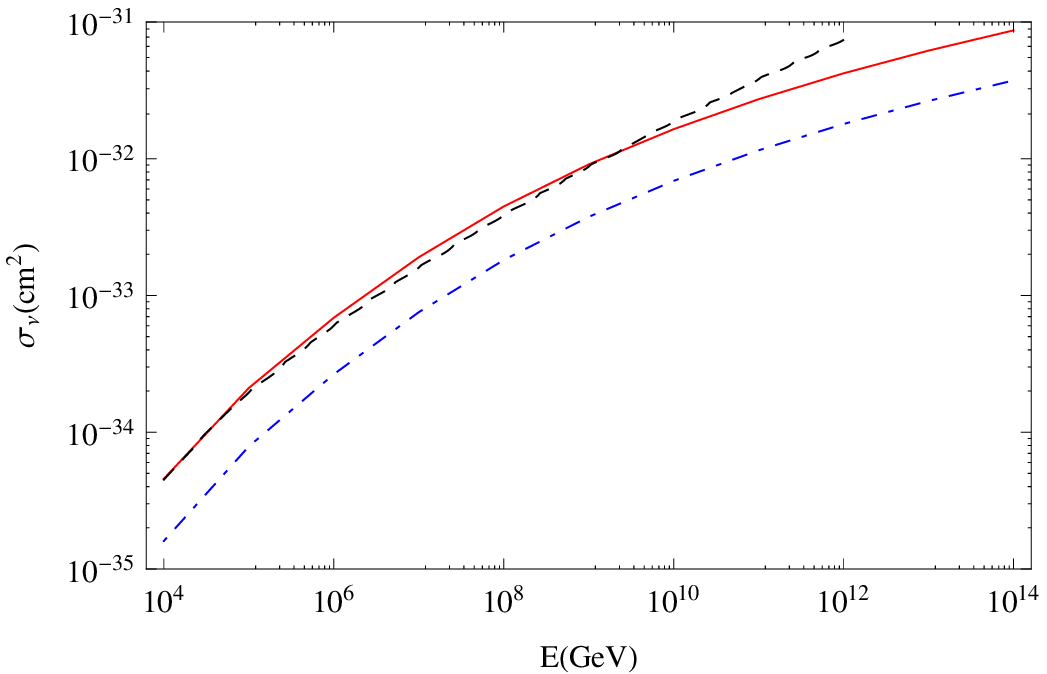} 
   \caption{This plot shows  the total charged current and neutral current cross sections for neutrino scattering on an isoscalar nucleon target, with $E_\nu$ in GeV and $\sigma_\nu$ in cm$^2$.  For charged currents,  HERA, our fit, is the (red) solid curve and the next-to-leading order C-SS  \cite{c-ss} fit is the (black) dashed curve.  For neutral currents, the (blue) lower  dot-dashed curve is for our new HERA results. Note that neutral currents are not included in the C-SS  study.}
   \label{fig:ccncfig2}
\end{figure}

The results are summarized graphically for the evaluations $\sigma_{HERA}$ and $\sigma_{C-SS}$  for CC, together with $\sigma_{HERA}$ for NC in Fig. \ref{fig:ccncfig2}.  Our new results are shown as a (red) solid upper curve for $\sigma_{CC}$ and a (blue) dot-dashed lower curve for $\sigma_{NC}$.  The perturbative QCD results of \cite{c-ss} for $\sigma_{CC}$ are shown by the (black) dashed curve up to $E_{\nu}$ = 10$^{12}$ GeV, the maximum energy they report.  The latter reference does not present NC results. Though they are not shown in Tables 2 and 3, we note in passing that the errors in our cross section values (including correlation terms) that are generated by the uncertainties in the fitted parameters range between 1\% and 1.7 \% in the interval from $10^5$ GeV to $10^{14}$ GeV, and are thus quite  small in comparison to the differences between our calculations and the pQCD calculations for energies above $10^9$ GeV, as illustrated in Fig. 1.
In both the CC and the NC cases, we see from the tables and from the plot that the pQCD extrapolation for $\sigma_{\nu N}(E)$ rises {\em significantly} above our "Froissart" extrapolation for $E_\nu> 10^9$ GeV.  

\section{Summary and Conclusions}
We have updated our previous prediction of the ultrahigh energy CC and NC neutrino cross sections by repeating our earlier calculation with a recently published, combined ZEUS and H1 data set with nearly double the statistics compared to that used in our calculation based on the 2001 ZEUS results alone.  Besides increasing the size of the data sample, the new analysis made a thorough study of the correlated errors, resolving the tension between the individual analyses. The results we report here use a 7 parameter fit, achieving a $\chi^2$/d.f. of 1.17, similar to that obtained by the previous 6 parameter fit to the much smaller ZEUS data set.  As indicated by the tabular and graphical results summarized in Sec. 3, the CC cross section predictions of pQCD and our model in the UHE range 10$^6$ $\leq E_{\nu} \leq$ 10$^8$ GeV agree reasonably well.  This is not too surprising, since the average $x$ and $Q^2$ values contributing to $\sigma_{\nu}(E_{\nu})$ in this range are both consistent with the "wee parton" model we invoke and lie  within the HERA data range. What is somewhat  surprising is  the  good agreement for $E_{\nu} < $10$^6$ GeV.

When $E_{\nu}$ $\geq$ 10$^9$ GeV, the $\ln^2 (1/x)$ term dominates the values of $F_2$, and the pQCD-based value of $\sigma_{\nu N}(E_{\nu})$, which is growing approximately as a fractional power of $x$, become clearly larger than our fit based on "Froissart-unitarity" growth with $E_{\nu}$. Note, however, that \cite{c-ss} finds that HERA I data---the input to the ZEUS-S global pdf fits \cite{zeus03}---indicate that  the growth of the cross section is not quite a power law and falls a bit below the earlier determination by Gandhi, Quigg, Reno and Sarcevic (GQRS) \cite{gqrs98} at energies above 10$^9$ GeV.  At the value $E_{\nu}$ = 10$^{12}$ GeV, for  example, the Ref. \cite{gqrs98} value for the CC cross section is a factor of 1.5 larger than that of Ref. \cite{c-ss}.  

Though the data set used for this work is much larger, basically doubling the statistics, and the error analysis more complete than the set used in our previous study \cite{bbmt}, there is qualitative and general quantitative agreement between the two at the very highest energies, even when extrapolated far above the kinematic region where the data lie.  We view this as  evidence for the stability and robustness of our extension of the data to ultrahigh energies.  

In conclusion, our results based on the complete HERA I data set confirm the message of our previous work that hints of unitarity constraints may already be present in the HERA  deep inelastic scattering evaluations of $F_2$. The unitarity constraints are \emph{most evident} in the predictions of the neutrino-nucleon total cross sections in the experimentally important realm of the highest energy cosmic rays, $E_{\nu} > $10$^9$ GeV, where cosmic ray experiments \cite{HiRes, Auger} and others such as RICE \cite{rice}, ANITA\cite{anita}, GLUE \cite{glue} and FORTE \cite{forte} have set limits on the fluxes of cosmic neutrinos.  The cross-section issue will ultimately be settled by an analysis such as that outlined in Ref. \cite{achs}. There the experimental requirements are laid out that would enable one to distinguish among the unscreened QCD (the pQCD approach), screened QCD, estimated in the "dipole" approximation (whose predictions resemble ours at the highest energies), and a radical non-linear picture like the quark-gluon condensate.  Meanwhile, in planning for new experiments and analysis of ongoing ones, our direct extrapolation from data, which does not use QCD evolution \cite{dglap} but builds in unitarity constraints, should play a critical role in estimating event rates and detection thresholds.

{\em Acknowledgments:} M. M. Block thanks the Aspen Center for Physics for its hospitality, P. Ha thanks Towson University Fisher College of Science and Mathematics for travel support and 
D. W. McKay received support from DOE Grant No. DE-FG02-04AR41308.

\end{document}